# Engineering Moiré Meta-crystals with Conventional Photonic and Phononic Structures

Mourad Oudich[1, 2], Xianghong Kong[3], Tan Zhang[3], Chengwei Qiu[3] and Yun Jing[1]

[1]*Graduate Program in Acoustics, Penn State University, University Park, Pennsylvania 16802, USA*

[2]*Université de Lorraine, CNRS, Institut Jean Lamour, F-54000 Nancy, France*

[3]*Department of Electrical and Computer Engineering, National University of Singapore, Singapore 117583, Singapore*

Corresponding authors: chengwei.qiu@nus.edu.sg , yqj5201@psu.edu

**Abstract**

Recent discoveries on Mott insulating and unconventional superconducting states in twisted bilayer graphene with Moiré superlattices have reshaped the landscape of "twistronics" and paved the way for developing high-temperature superconductors and new devices for quantum computing and sensing. Meanwhile, artificially structured photonic and phononic metamaterials/crystals (or meta-crystals) have become a fertile playground for emulating quantum-mechanical features of condensed matter systems, revealing new routes for robust control of classical waves. Drawing inspiration from the success of twisted bilayer graphene, this perspective casts an overarching framework of the emerging Moiré photonic and phononic meta-crystals that promise novel classical-wave devices. We begin with the fundamentals of Moiré superlattices, before highlighting recent works that exploit twist angle and interlayer coupling as new ingredients to engineer and tailor the band structures and effective material properties of photonic and phononic meta-crystals. We finally discuss future directions and promises of this emerging area in materials science and wave physics.

**Main**

Van der Waals (vdW) heterostructures are two-dimensional (2D) atomic layer heterostructures where the interlayer binding is achieved via weak vdW interactions[1]. The study on how relative twist angle between successive layers in vdW heterostructures can be used to manipulate the material's electronic properties is often referred to as twistronics[2]. An increasingly important topic in twistronics is twisted bilayer graphene (TBG), where two graphene sheets are placed on top of each other with a slight angle misalignment[3]. Such a small twist results in an artistic Moiré superlattice at a much larger length scale than the underlying graphene lattice (**Fig. 1a**), radically changing the band structure of bilayer graphene with the more conventional AA-stack and AB-stack (Bernal) configurations, which in turn gives rise to unconventional electronic[4,5], optical[6,7], and thermal properties[8] of TBG. One of the most extraordinary features of TBG is the emergence of zero-energy-level flat bands at a series of so-called magic angles.

In 2011, it was first predicted by Bistritzer and MacDonald that the Dirac-point velocity vanishes at some magic angles (the smallest being around 1.05°)[5]. Nearly flat bands would emerge at the magic angle, which contributes a sharp peak to the Dirac-point density of states (DOS) (**Fig. 1b**). This study marked an important milestone for the theoretical work in twistronics. It was not until 2018, that the magic-angle bilayer graphene was experimentally confirmed by Jarillo-Herrero's group at MIT (**Fig 1c**). Their back-to-back papers on Nature reported on two ground-breaking discoveries pertaining to the magic-angle bilayer graphene: correlated (Mott) insulation[9] and unconventional superconductivity at around 1.7 K[10] (**Fig. 1d**). These two discoveries have generated a host of theoretical and experimental papers seeking to better understand and further explore exotic phenomena associated with magic-angle TBG[11–15].

Taking inspiration from TBG, researchers in classical waves have attempted to employ the twist degree of freedom as a new dimension to expand the design space of synthetic photonic and phononic meta-crystals, such as photonic crystals (PtC) and phononic crystals (PnC). For example, stacking up two PtCs or PnCs in honeycomb lattice with a small twist angle between the two layers constitutes a simple analogue of TBG[16–21]. The interlayer coupling, in this case, is typically provided by the medium in between the two layers, where evanescent wave fields from each monolayer are coupled. Using these bilayer Moiré structures, magic angles have been numerically observed for electromagnetic and mechanical waves, revealing a new route for flat band engineering in synthetic photonic and phononic meta-crystals. Beyond mimicking magic angle bilayer graphene, some of the Moiré photonic and phononic bilayer designs enable the interlayer coupling strength, which is extremely difficult to tune in TBG, to become a new degree of freedom to tailor the meta-crystal's response to waves[16,18]. This new feature of classical waves unlocks the access to an uncharted domain with ultra-strong interlayer coupling and large magic angles (>3°) unattainable in TBG. This new degree of freedom of interlayer coupling provides a strong ground for exploring new physics arising from features uniquely associated with Moiré photonic and phononic meta-crystals, so as to move beyond simple emulation of behaviors observed in TBG. Another example in this regard is the demonstration of polaritonic topological transition in bilayers of α-phase molybdenum trioxide (α-MoO$_3$)[22], where the polariton dispersion can be precisely controlled by the twist angle and shows a transition point at a fundamentally different magic angle from the one in TBG. This work exemplifies how researchers can draw inspiration from TBG to introduce new design paradigms to enable optical and acoustic properties unavailable in monolayer synthetic photonic and phononic meta-crystals.

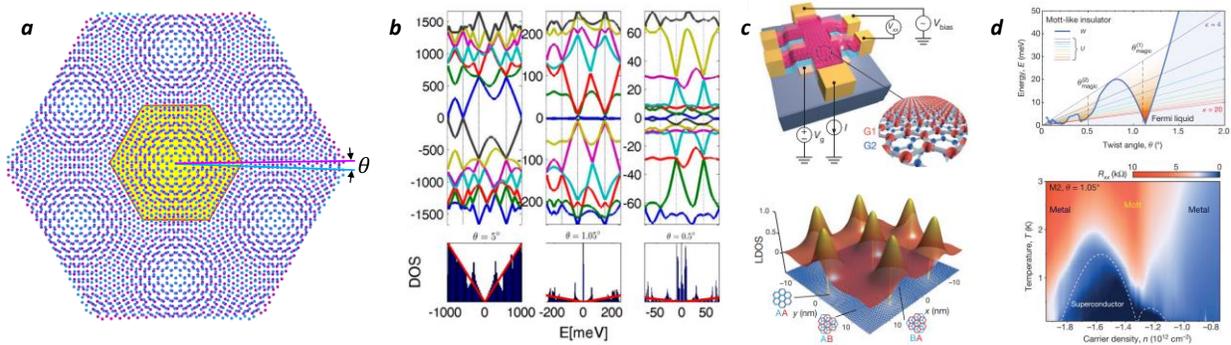

**Fig 1. | Twisted bilayer graphene.** *a*, Stacking up two graphene sheets with a twist in between leads to a Moiré superlattice. The unit cell is highlighted with a yellow hexagon. *b*, Electronic energy dispersion of the TBG for twist angles of θ=5° (left), θ=1.05° (middle), and θ=0.5° (right). Flat bands emerge at around 1.05°, one of the magic angles of TBG. The corresponding DOS plots are shown on the bottom panel[5]. *c*, Experimental characterization of TBG for superconductivity[10]. The bottom panel shows the local DOS associated with the flat band at the magic angles[9]. *d*, Bandwidth W of the flat-band branch in TBG and the on-site energy as a function of the twist angle θ. Near the magic angles, the TBG becomes a Mott-like insulator[9]. The bottom panel shows the measured resistance as a function of temperature at a magic angle, showing a superconductivity dome[10].

In this perspective article, we will showcase recent developments in the field of Moiré classical-wave meta-crystals. Specifically, we will begin by introducing the Moiré pattern and its underlying mathematics. We will then examine Moiré photonic meta-crystals, followed by Moiré phononic meta-crystals. Our discussion will delve into the construction of these bilayer structures, the characterization of their wave dispersion behavior, and the customization of their twist angle and interlayer coupling strength to realize unique wave phenomena for both electromagnetic and mechanical waves. We will also address the challenges and routes for applications in this field, and conclude by highlighting the future directions for the development of photonic and phononic Moiré meta-crystals. We would like to highlight that there is a recent review paper on Moiré photonics and optoelectronics[23]. Our paper, in turn, offers a complementary perspective to further illuminate the emerging field of Moiré physics.

**The Moiré pattern**

The Moiré pattern is a well-known visual effect that can be observed when at least two planar periodic structures such as grids or gratings are superposed or overlaid closely to each other with a misalignment, such as an angular twist. The Moiré pattern is often seen as an alternation of dark and bright areas, where the bright areas are created when the two layers largely overlap, while the dark areas are ones where the two layers overlap to a much lesser extent. The Moiré pattern is highly sensitive to geometrical displacement and rotation of one layer with respect to the other, which renders the Moiré-inspired system an extremely dynamic platform for a wide range of applications. For instance, Moiré patterns have been used in strain distribution measurements for the prediction of cracks[24,25], metrology for the detection of angle variations[26], precise alignment sensing[27,28], and for displacement[29] and movement[30] measurements. Interestingly, Moiré patterns have also been historically used for marine navigation in shoreside beacons called "Inogon lights" to indicate safe paths for ships and prevent them from running into underwater cables and pipelines.

In condensed matter physics, Moiré patterns are typically created from angular misalignment between a pair of atomic monolayers (**Fig. 1a**). Of particular interest is the TBG, largely owing to the zero-energy level flat bands at the magic angles. Here, we delineate the geometrical construction of the Moiré superlattice from twisting two monolayers of atoms. Let the periodicity of the first lattice be governed by two primitive lattice vectors $a_1$ and $a_2$. The position of each unit cell (atom) can be then described by the vector $R_a = n_1 a_1 + n_2 a_2$ where $n_1$ and $n_2$ are integers.

Assuming the second lattice is identical to the first one, but rotated by an angle $\theta$, each unit cell in the second lattice would be located at $\boldsymbol{R}_{a'} = m_1 \boldsymbol{a}'_1 + m_2 \boldsymbol{a}'_2$, where $m_1$ and $m_2$ are integers and $\boldsymbol{a}'_1$ and $\boldsymbol{a}'_2$ are vectors after rotation of $\boldsymbol{a}_1$ and $\boldsymbol{a}_2$. In the $(x, y)$ plane, if we adopt the complex notation $a_k = a_{k,x} + j a_{k,y}$ where $a_{k,x}$ and $a_{k,y}$ are the coordinates of the vector $\boldsymbol{a}_k$ ($k = 1; 2$), then one can write $a'_k = a_k e^{j\theta}$, which leads to $R_{a'} = m_1 a'_1 + m_2 a'_2 = (m_1 a_1 + m_2 a_2) e^{j\theta}$. In general, the equality $R_{a'} = R_a$ is not necessarily valid for any set of integers $n_1, n_2, m_1$, and $m_2$, meaning that the resulting Moiré patterns are not guaranteed to be perfectly periodic — they can be instead quasiperiodic. These twist-induced Moiré patterns, however, can become perfectly periodic for discrete values of $\theta$ which can be determined by the set of $n_1, n_2, m_1$ and $m_2$ that lead to $R_{a'} = R_a$. These angles, known as commensurate angles, can be given by Eq. (1), as long as the solution is a real number[31],

$$\theta = i \ln \left( \frac{n_1 a_{1,x} + n_2 a_{2,x} + i(n_1 a_{1,y} + n_2 a_{2,y})}{m_1 a_{1,x} + m_2 a_{2,x} + i(m_1 a_{1,y} + m_2 a_{2,y})} \right) \quad (1)$$

A trivial solution would be $m_1 = n_1$ and $m_2 = n_2$, which corresponds to $\theta = 0$.

For the case of triangular, honeycomb, and Kagome lattices, the periodicity can be defined by the set of vectors $\boldsymbol{a}_1 = a(1,0)$ and $\boldsymbol{a}_2 = a(1/2, \sqrt{3}/2)$ where $a$ is equal to $p$, $\sqrt{3}p$, or $2p$ for the triangular, honeycomb, and Kagome lattice, respectively, with $p$ being the distance between closest atoms. Subsequently, Eq. (1) becomes,

$$\theta = i \ln \left( \frac{2n_1 + n_2 + i n_2 \sqrt{3}}{2m_1 + m_2 + i m_2 \sqrt{3}} \right) \quad (2)$$

However, in the case of twisted bilayer square lattices, Eq. (1) becomes,

$$\theta = i \ln \left( \frac{n_1 + i n_2}{m_1 + i m_2} \right) \quad (3)$$

Then, $\theta$ being real valued requires $m_1^2 + m_2^2 + m_1 m_2 = n_1^2 + n_2^2 + n_1 n_2$ for triangular, honeycomb, and Kagome lattices, while this relation becomes $m_1^2 + m_2^2 = n_1^2 + n_2^2$ for the case of the square lattice. Apart from the trivial solution, one can consider the solution $m_1 = n_2$ and $m_2 = n_1$ that gives the commensurate angles of the twisted bilayer lattices. **Figure 2** shows the discrete commensurate angles as a function of these integers for twisted bilayers with triangular, honeycomb, and Kagome lattices (**Fig. 2a**), and the square lattice (**Fig. 2b**). These figures suggest that the distribution of commensurate angles in the case of triangular, hexagonal, or Kagome lattices are slightly denser in comparison to that of the twisted bilayer square lattice. **Figure 2c** also shows examples of unit cells of the Moiré superlattices for one small and one large commensurate twist angles. The number of atoms per unit cell is greater in the case of twisted bilayer Kagome lattice than honeycomb and triangular ones. Note that the superlattices are drawn in a way that they have the identical size for different cases, resulting in atoms at different sizes.

Among these four types of twisted bilayer lattices, the most frequently studied one is the TBG (honeycomb lattice) with the exploration of its electronic properties[5,9–11,13,32], optical[6,7,33–35], and thermal properties[8,36,37]. Analogs of TBG in optics, acoustics, and elastodynamics have been devised for demonstrating wave confinement and flat band dispersion at the magic angles as well as other interesting behaviors, which will be discussed in the subsequent sections. Meanwhile, fewer works have explored the electronic dispersion of twisted bilayer triangular lattices using WSe2[38], MoSe2/WSe2[39], WSe2/WS2[40] and transition metal dichalcogenides[41], while other studies have investigated twisted bilayer Kagome lattices[42,43] and square lattices[44–47].

Moiré patterns can also be created by considering two lattices with mismatch in their periodicities along a specific direction. Consider a one-dimensional (1D) periodic lattice with a period of $a$ and a second lattice with a period of $a' = a + \delta a$ where $0 < \delta < 1$, then stacking the two lattices would yield a 1D Moiré pattern. This pattern is generally quasiperiodic but can also be perfectly periodic for discrete values of $\delta$, which can be found by using the equation $na = ma'$, where $n$ and $m$ are integers. This leads to the relation $\delta = (n - m)/m$ with the condition of $1 < n/m < 2$. **Figure 2c** shows the values of $\delta$ as a function of $n$ and $m$, while **Fig. 2d** presents examples of a 1D Moiré superlattice created from two 1D lattices and a 1D Moiré superlattice created from two layers of triangular lattices with a periodicity mismatch in the horizontal direction. These bilayer 1D Moiré patterns in classical waves have been studied less extensively than 2D Moiré patterns and the relevant works have been mainly focused on optics[48,49]. Finally, the mismatch in periodicity can be also created in both spatial directions[50].

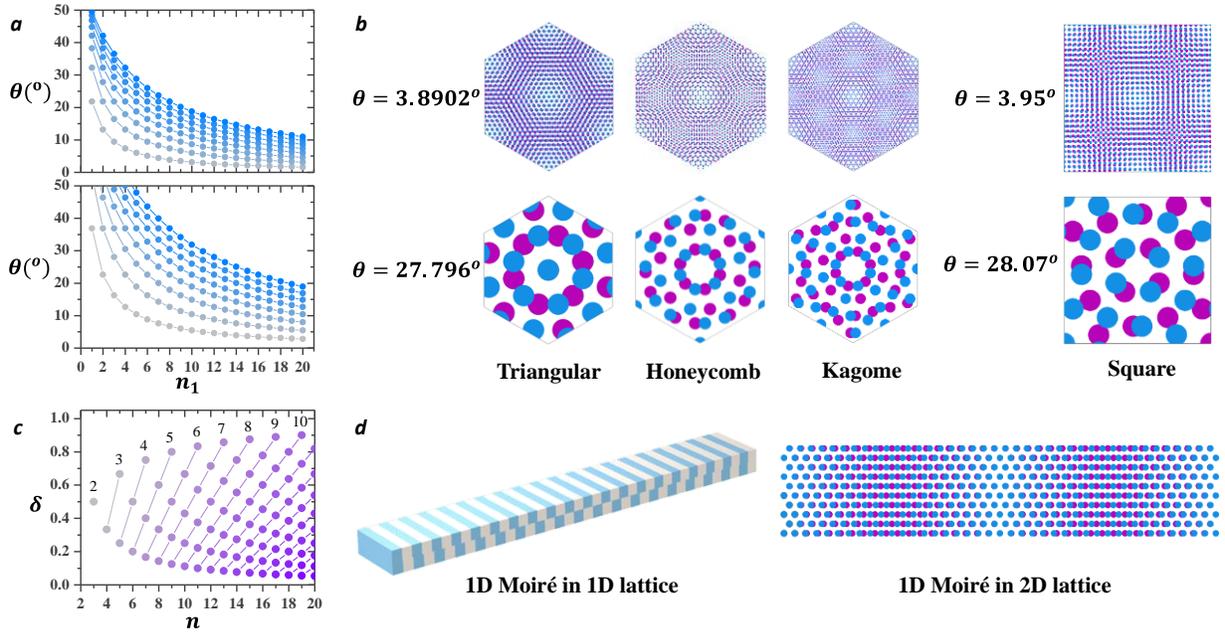

**Fig 2. | Construction of Moiré lattices. *a*,** Commensurate angles for twisted bilayer triangular, hexagonal, and Kagome lattices (upper panel) and square lattice (bottom panel) predicted from Eqn. (2) and (3). Note that twisted bilayer triangular, hexagonal, and Kagome lattices share the same commensurate angles. The horizontal axis represents the integer $n_1$ while each line of connected dots corresponds to a value of $n_2$ that varies from $n_1 + 1$ (gray) to $n_1 + 8$ (blue). ***b***, Examples of 2D Moiré patterns created by these twisted bilayer lattices for large and small commensurate angles. The atoms are drawn in different sizes so that the unit cells maintain the same size for different lattices. *c*, A set of periodicity mismatch rate $\delta$ that gives rise to perfect periodic 1D Moiré patterns. The number indicated for each line of connected dots is the value of the integer $m$. *e*, Examples of 1D Moiré patterns constructed by stacking up two 1D lattices or 2D lattices with periodicity mismatch in one direction.

## Moiré photonic meta-crystals

A Moiré photonic (phononic) meta-crystal is generally created by stacking two layers of photonic (phononic) lattices in a way that enables the interaction between the waves supported by each PtC (PnC) layer. This interaction can be generally controlled by the separation distance or the coupling medium between the layers, while the photonic (phononic) dispersion of the Moiré meta-crystal is further tailored via the twist of one photonic (phononic) layer with respect to the other. These new degrees of freedom expand the design space of photonic (phononic) lattices, leading to the development of novel twisted photonic (phononic) meta-crystals along with new theoretical/numerical approaches to describe their wave behaviors.

The generalized concept of homogeneous crystals is considered valid when the behavior of a crystal can be modeled by the effective medium theory. For photonics (or phononics), such a concept goes hand in hand with the long wavelength approximation, under which the crystal can be greatly simplified by taking its average electromagnetic (or mechanical) response — a process known as homogenization. For instance, certain natural crystals (e.g., black phosphorous[51], α-MoO$_3$[22,52,53], and WTe$_2$[54]) and nanostructured crystals with deep subwavelength periodicity (e.g., graphene and hBN nanoribbon arrays[55,56]) can be treated as homogeneous crystals, where an anisotropic surface conductivity tensor can be used to characterize these low-dimensional materials for manipulating light[7]. Owing to dispersion, the imaginary part of the surface conductivity tensor can have opposite signs in the two orthogonal directions within a certain frequency range, where the crystal would behave like a strongly anisotropic 2D material known as hyperbolic metasurface. When two such hyperbolic metasurfaces are stacked up, a Moiré metasurface is created (**Fig. 3a**). The isofrequency contour of the Moiré metasurface lies somewhere between those of the two individual metasurfaces due to the coupling effect (**Fig. 3b**). The isofrequency contour can be analytically derived by first choosing plane waves as the basis for the three domains created by the two metasurfaces (the domain above the top metasurface; the domain in between; the domain below the bottom metasurface), and then matching the boundary conditions of the two metasurfaces which are determined by the surface conductivity tensor. By tuning the twist angle between the two hyperbolic metasurfaces, the isofrequency contour evolves from being hyperbolic to elliptical, which is analogous to the Lifshitz transition in electronics[22,57,58]. The isofrequency contour flattens at a photonic magic angle where the transition from hyperbolic to elliptical contour occurs. The self-collimation phenomenon can be detected at the magic angle since the group velocity direction is fixed on the flattened isofrequency contour (**Fig. 3c**)[22,59]. In addition to the electrical surface conductivity, chiral surface conductivity was also discovered in photonic twisted bilayer graphene metasurfaces when retrieving the effective electromagnetic parameters[60–63]. The opposite chirality can be simply created by the relative rotation of the two layers which has mirror symmetry. Note that though graphene can be also considered an optical metasurface, the optical properties of TBG are beyond the scope of this perspective and the reader is referred to a recent review paper for more in-depth discussion on this topic[64].

When the working wavelength is comparable with the unit cell of the crystal, homogenization is no longer valid, and the Bloch's theorem can be applied to analyze the PtC's electromagnetic response. The periodicity of the Moiré pattern at a commensurate angle is usually much larger than the size of the unit cell of the monolayer. However, in the case of honeycomb lattice, when the twist angle is 0° (AA stacking) or 60° (AB stacking), the periodicity of the Moiré pattern reaches its minimum, which is the same as that of the monolayer (**Fig. 3d**)[18]. Additionally, narrow solitons appear between AB and BA domains and high local optical conductivity can be observed at the AA domains by the nano-imaging experiment[6,65]. By creating effective potential wells centered around AA stacked region, the intrinsic localized states are obtained, which leads to the superflat bands in a wide and continuous parameter space[66]. Hence, a detailed investigation of AA and AB stacking is crucial for understanding the underlying physics of Moiré patterns[18]. **Figure 3d** illustrates an example wherein a PtC layer comprises a metallic plate featuring a hexagonal lattice of metallic pillars that facilitate the propagation of spoof surface plasmons (SSP)[67]. The band structure of the SSP resulted from such a lattice mimics that of the graphene, with a Dirac cone appearing at the $K$ point of the reciprocal lattice[68]. The stacking of two of these PtC lattices at an appropriate separation distance allows the SSPs from the two layers to interact, leading to a dispersion behavior that is strikingly similar to that of the bilayer graphene[18]. Furthermore, the tight binding model developed in bilayer graphene can be readily used to describe the dispersion of the bilayer PtC at the vicinity of the Dirac frequency with properly fitted parameters[18] (**Fig. 3d**).

Besides, different quantitative analyses have been applied to the photonic analogy of TBG[21,69]. Dong *et al.* used a silicon disk as the photonic counterpart of the carbon atom in graphene (**Fig. 3e**)[21]. The coupled mode theory was applied to describe the coupling between nearest-neighbor disks. For simplicity, a continuum model of the interlayer coupling strength was considered to replace the discrete coupling between two disks. Compared with the continuum model for the homogeneous crystals[58] mentioned above, the interlayer coupling strength in the coupled mode theory is periodic and has the same periodicity as the Moiré pattern. The band diagram and DOS calculated from the continuum model show local flat bands at the photonic magic angles (**Fig. 3e**). Similar photonic properties of twisted bilayer photonic honeycomb lattices have also been demonstrated by Tang *et al.*[20] , where the band structure was engineered by adjusting the device geometry and a larger band asymmetry was shown in the photonic system. While these two studies[20,21] numerically demonstrated the magic angles in twisted bilayer photonic graphene at the optical frequency, Oudich *et al.*[18] demonstrated the magic angles as well as topological corner modes in a twisted bilayer photonic graphene at the microwave frequency. Further, these studies[18,20,21] explored interlayer coupling strength as a degree of freedom to tune the magic angle. Besides the prestigious magic angle flat band hosted by Moiré structures, other exotic optical states can also be approached by tuning the angle between two photonic graphenes. Very recently, Huang et al.[70] theoretically demonstrated quasi-bound states in the continuum in a Moiré PtC at the THz frequency. In addition to the honeycomb lattice, Lou et al.[69,71] investigated the twisted bilayer square lattice photonic slab through a high-dimensional plane wave expansion method (**Fig. 3f**). Instead of choosing plane waves as the basis as mentioned in previous effective medium theory, Bloch waves were first chosen as the basis of the two slabs and plane waves were used as the basis of the surrounding space. Bloch waves in the slabs were then decomposed into plane waves and boundary conditions with the plane waves in the surrounding spaces were matched to collectively give rise to the analytical solution. Strongly tunable resonance properties and chiral behavior were discovered by observing the transmission under incident light with different frequencies and twist angles. The same type of twisted bilayer PtC slabs were also demonstrated to be a tunable narrow stop band frequency filter[47]. Interestingly, topological flat bands could be sustained in Moiré photonic materials, where topological edge modes would deform into higher-order topological corner modes after breaking the reflection symmetry of the boundary of the superlattices[72].

When the twist angle is incommensurate, the photonic Moiré pattern becomes aperiodic without translational periodicity while the rotational symmetry still persists[45,73–75]. Instead of using a bilayer system to generate the Moiré pattern, Wang et al.[45] projected the Moiré patterns onto a single surface using optical induction and the weight of the "two layers" can be tuned during the projection process to generate different Moiré patterns. They experimentally demonstrated the localization-delocalization transition of light by altering the patterns from incommensurable to commensurable. As a particular case of the incommensurate lattice, the quasilattice refers to the case when the lattice vectors have the equiangular offset between them and are of equal magnitude[73]. The 45° twist angle in a square lattice or 30° twist angle in a hexagonal lattice can form a quasilattice, which has 8-fold and 12-fold rotational symmetry, respectively (**Fig. 3h**). Quasilattice patterns with rotational symmetries as high as 36-fold were developed by Moiré nanolithography on silver plasmonic crystals[74], and an increased number of surface-plasmon-polariton modes have been discovered in quasilattice[74]. In a recent study, Zhang et al.[76] proposed a theoretical approach based on combining supercell calculation and band unfolding techniques to globally characterize the photonic dispersion of a 2D quasiperiodic Moiré superlattice. Compared with the typical near-field Moiré photonic meta-crystals, most recently, far-field coupling between Moiré photonic architectures have been experimentally observed, where twist-angle-controlled directional lasing emissions were achieved[77].

In addition to the twist between two monolayer crystals, Moiré patterns due to mismatched lattice constants have also been studied[48,49,78]. For instance, two parallel hexagonal lattice metallic ring metasurfaces with a lattice constant mismatch in one direction were introduced to form a Moiré bilayer system (**Fig. 3g**)[48]. Since the periodicity of the Moiré pattern is much larger than the unit cell, the supercell was decomposed into unit cells with different shifts between the two layers and the relative shift in the unit cell was treated as an effective gauge field created by an artificial magnetic field. The corresponding photonic Landau levels were observed experimentally. Similar results can also be found when overlapping two 1D PtC slabs with mismatched periods[49,78,79] where the authors showed a high concentration of the Wannier function in a Moiré cell. In a recent study, Talukdar *et al.*[80] designed and fabricated a 1D Moiré silicon photonic nanowire to demonstrate a host of behaviors including slow-light, high Q-factor moiré resonators, multi-resonant filters, suppression of grating sidebands, persistent vs extinguishable transmission, tunable Q-factors, and tunable group velocities.

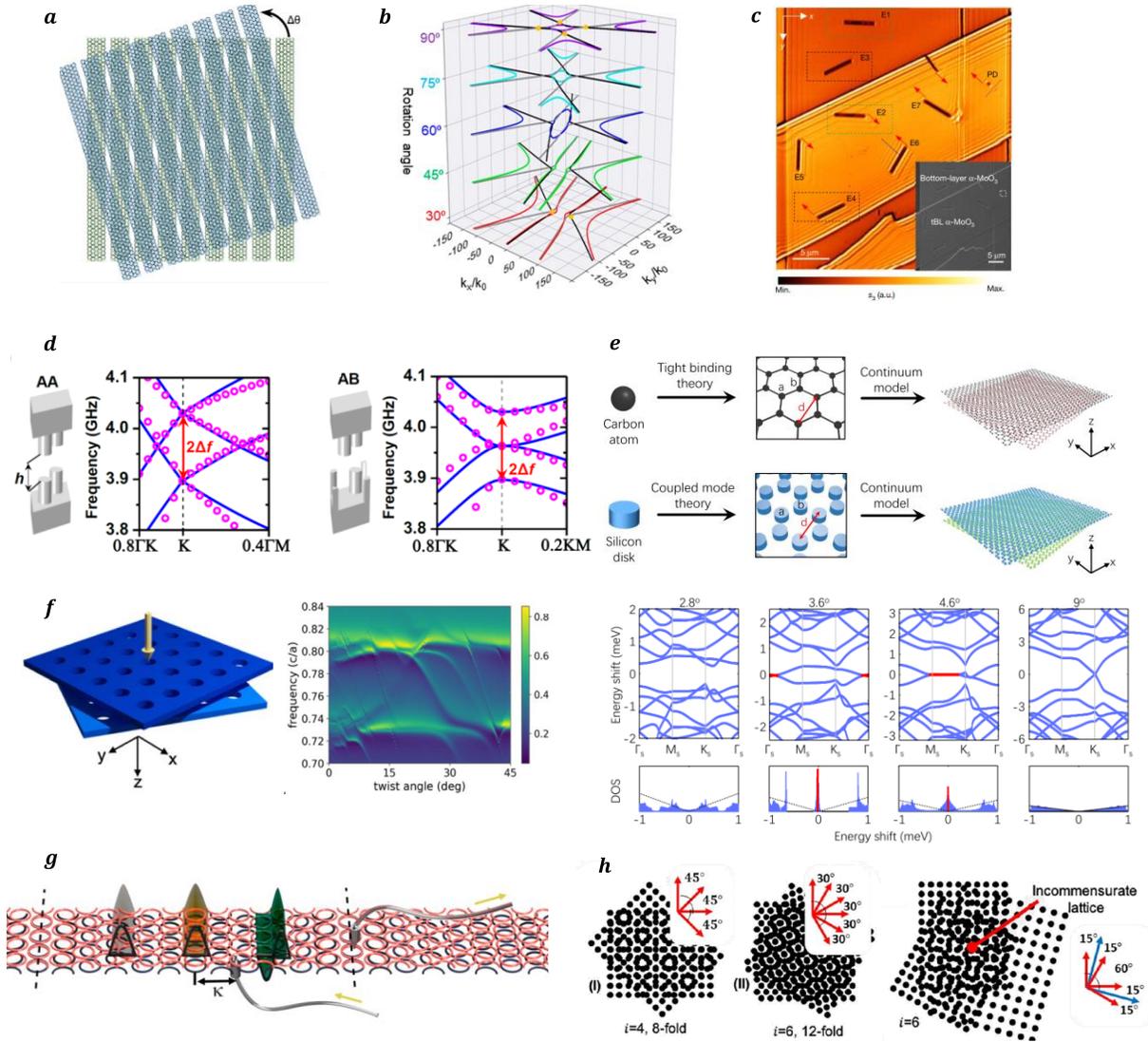

**Fig 3. | Examples of Moiré photonic meta-crystals. a,** The Moiré hyperbolic metasurface composed of two coupled uniaxial metasurfaces with a certain twist angle[56]. **b,** Dispersion relations as a function of the twist angle. The gray (black) solid lines denote the dispersion of the first (second) individual metasurface while the colored lines denote the dispersion of the Moiré metasurface[56]. **c,** Near-field images of fixed group velocity direction (red arrows) near the magic angle[22]. **d,** AA stacked and AB stacked bilayer photonic crystals and their associated band structures[18]. **e,** Twisted bilayer photonic crystal based on a honeycomb lattice of silicon nanodisks and its comparison with the TBG. The dispersion curves and DOS with different twist angles show the existence of local flat bands at certain angles[21]. **f,** Twisted bilayer photonic crystal with circular holes in a square lattice and the transmission as a function of frequency and twist angle[69]. **g,** Moiré pattern formed using metallic rings with different periods in the top and bottom layers[48]. **h,** 8-fold quasilattice, 12-fold quasilattice, and a general incommensurate lattice formed by the overlap of square lattices from two layers[73].

**Moiré phononic meta-crystals**

TBG not only sparked a substantial interest in developing Moiré photonic meta-crystals, but also spawned the new field of Moiré phononic meta-crystals for controlling acoustic[16,19,81–84] and elastic waves[17,31,85–87]. Lu et al.[81] proposed a sonic bilayer structure composed of two stacked phononic crystals (PnCs), where each monolayer PnC is made of a triangular lattice of rigid triangular units (**Fig. 4a**). A perforated rigid plate separated the two PnCs, where the holes induce the interlayer coupling. The stacking was of AA configuration while the triangular units within each cell were rotated in both layers to give rise to different dispersion behaviors at the vicinity of the Dirac point. The existence of two types of topological valley edge states was numerically and experimentally demonstrated, with interfaces that support either valley Hall states propagating in both layers or layer-valley Hall states that mainly propagate in a single layer. Topological waveguiding with propagation from one layer to the other was also experimentally demonstrated (**Fig. 4a** lower panel). Though this work does not directly involve Moiré patterns, it is one of the earliest studies that provide crucial insight on how interlayer coupling and rotation can be harnessed to engineer the dispersion of bilayer PnCs. Dorrell et al.[82] proposed a bilayer PnC consisting of rigid cylindrical rods in honeycomb lattice and the two layers were separated by a thin vibrating membrane to ensure the interlayer coupling of acoustic waves (**Fig. 4b**). By choosing the proper interlayer coupling strength via changing the thickness and the density of the membrane, the authors numerically showed that the acoustic dispersion of the bilayer can mimic the electronic dispersion of the classical bilayer graphene near the Dirac cone frequency for both AA and AB stacking configurations with two sets of crossing Dirac bands and quadratic dispersion, respectively (**Fig. 4b** lower panels). Shortly after, the twist degree of freedom was considered by Deng et al.[16] who introduced a sonic bilayer crystal where each layer is a rigid plate with a honeycomb lattice of cylindrical air cavities (**Fig. 4c**). Each PnC plate supports spoof surface acoustic waves (SSAW) propagating in the near field above the air cavities with evanescent decay in the direction perpendicular to the plate surface. By positioning the phononic plates to face each other with an air gap in between, the SSAW supported by each plate can interact, mirroring the interlayer hopping in bilayer graphene. Further, twisting one plate with respect to the other creates a Moiré pattern, and it was numerically shown that at specific twist angles (magic angles), flat bands appear with confined acoustic intensity in the AA regions of the Moiré superlattice (**Fig. 4c** lower panel). The magic angle strongly depends on the interlayer coupling strength and can be tuned by varying the distance between the PnCs. Gardezi et al.[19] designed a bilayer twisted acoustic metamaterial using a vibrating polyethylene membrane as the coupling medium (**Fig. 4d**). The authors also numerically showed trapping of sound via the twist which is associated with the flattening of the Dirac bands at a magic angle of 1.12º (**Fig. 4d**). This magic angle can be tuned to higher values by changing the interlayer coupling strength through varying the thickness of the membrane. The dispersion of these bilayer PtCs can be described at the vicinity of the Dirac frequency by formulating the Hamiltonian from the tight binding model of the bilayer graphene. Recently, Wu et al.[83] presented an acoustic bilayer design at a large twist angle of 27.79º, consisting of connected cavities. Strong interlayer coupling was used to generate a band gap that harbors higher order topological states.

In the context of elastodynamics, Rosendo López et al.[17] theoretically investigated the analog of TBG for elastic waves by two weakly coupled vibrating plates via a thin elastic medium, where each plate is attached with a honeycomb lattice of point masses (**Fig. 4e**). The underlying physics of the interlayer coupling is the interaction of flexural waves hosted by the plates. The authors developed a theoretical model to describe their system based on Germain–Lagrange approximation from the equation of motion governing flexural waves in coupled plates. The twist angle comes into play when describing the mass distribution on both layers. They demonstrated the emergence of flat bands at a magic angle of 1.61º (lower panel of **Fig. 4e**). Meanwhile, Martí-Sabaté and Torrent[31] conducted a theoretical study on the interaction of elastodynamic modes with a cluster of scatterers distributed in a Moiré pattern over a thin plate. This study revealed the emergence of dipolar resonances at specific discrete values of the "twist angles" (**Fig. 4f**). Yves et al.[86] constructed a plate decorated with a lattice of pillars with modulated heights in a Moiré pattern, and demonstrated topological transition of the isofrequency contour from hyperbolic to elliptical dispersion, similar to what was observed in a previous work in photonics[22]. Very recently, Oudich et al.[87] presented a family of bilayer PnC, where both sides of a plate are decorated with a hexagonal lattice of pillars. A plate with a sufficiently large thickness possesses a weak interlayer coupling between surface acoustic waves (SAW) propagating on each side of the plate, representing a direct analogue of bilayer graphene. The authors also studied the twisted bilayer PnC under a large commensurate angle of 38.213º, which creates a structure with an even sublattice exchange (SE) symmetry. Furthermore, by lowering the thickness of the plate, strong interlayer coupling can be introduced which leads to substantial changes of the band structure and the possibility of bilayer Valley Hall states under the even SE symmetry.

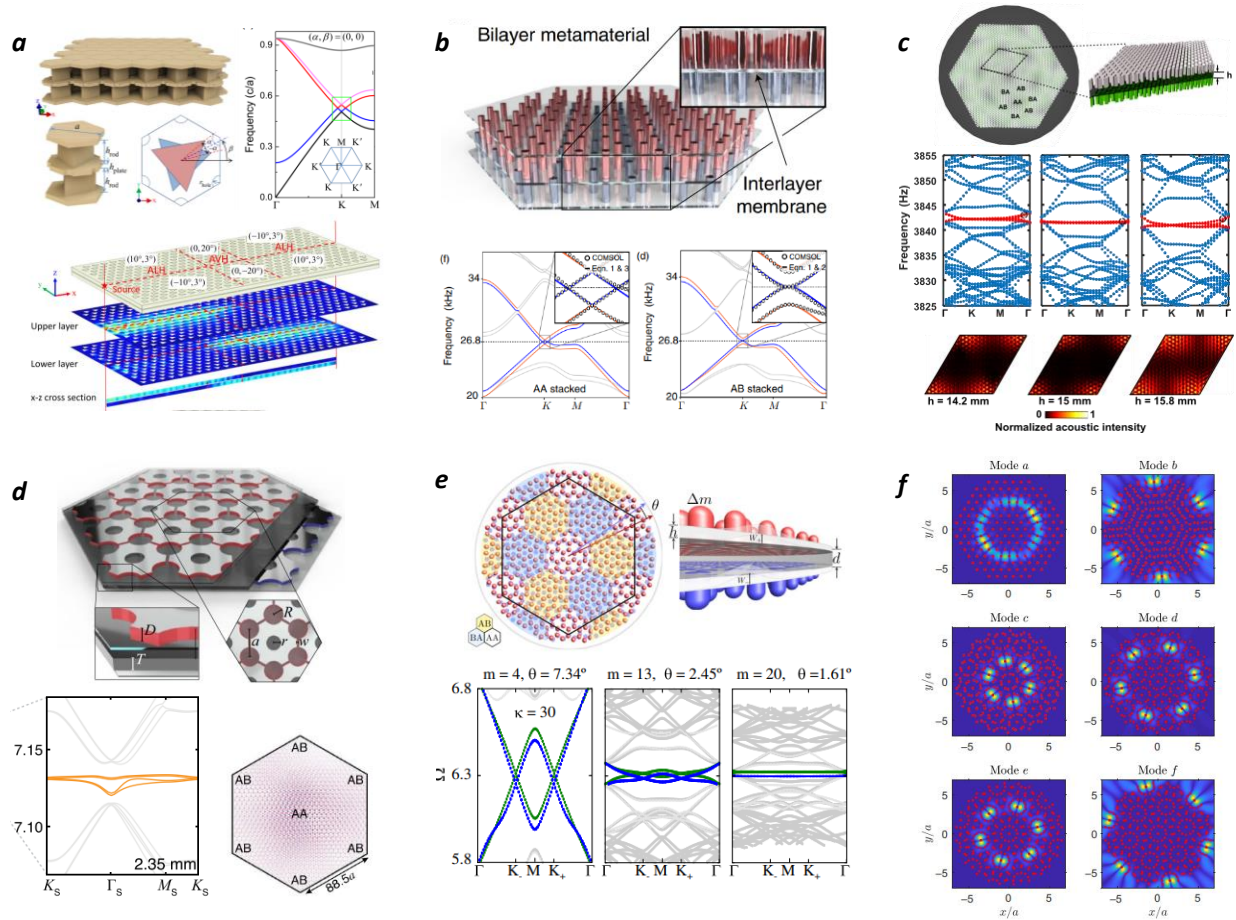

**Fig 4. | Examples of Moiré phononic meta-crystals.** *a*, Bilayer sonic crystal made of two PnCs with triangular lattices, and a rigid plate with holes separates the two PnCs to enable mode coupling[81]. (Right panel) The band dispersion shows crossing bands for AA staking. (Lower panel) Bilayer valley Hall transport from the upper layer to the bottom one. *b*, Bilayer metamaterial made of two acoustic lattices of rigid cylinders in air separated by a thin vibrating membrane to mimic the interlayer hoping[82]. (Lower panel) band dispersions of AA and AB stacking. *c*, Twisted bilayer sonic crystal consisting of two rigid plates with air cavities facing each other, where each plate supports the propagation of SSAW[16]. (Lower panel) band structures at a fixed twist angle of 3.481° for different interlayer couplings via changing the airgap thickness between the rigid plates. The acoustic intensity in the Moiré supercell is plotted near the Γ point, for the bands marked in red color. *d*, Acoustic bilayer structure created from stacking two sonic crystals, each made of connected cavities. The sonic crystals are separated by a vibrating membrane[19]. (Lower panel) flat bands created at the magic angle with confined acoustic energy at the AA regions of the Moiré supercell. *e*, Twisted elastic bilayer lattice made of two coupled plates decorated with honeycomb lattices of pillars, giving rise to flat bands at the magic angle[17]. *f*, Moiré lattices made of a cluster of scatterers displaying dipolar resonances at discrete values of twist angle[31].

**Outlook**

As it currently stands, there are two different research directions for Moiré photonic and phononic meta-crystals. The first one is centered on identifying engineered artificial materials that control waves to emulate the electronic behaviors experimentally observed or theoretically predicted in TBG. The epitome of this effort is the finding of flat bands at magic angles in bilayer photonic[18,21], sonic[16,19], and elastodynamic[17] graphene. While it is intriguing to show that the concept of magic angle can be generalized to virtually all classical wave systems, this twist-induced behavior (flat bands at the magic angle) has yet to be experimentally observed in a classical wave counterpart of TBG. This is largely because at small twist angles, the unit cell becomes extremely large, especially for Moiré phononic meta-crystals and microwave Moiré photonic meta-crystals due to the large wavelength that is often employed, and therefore the fabrication of a full macroscopic sample comprising a sufficient number of unit cells becomes impractical. Even if such a sample can be built, experimental characterization can be exceedingly challenging. This is because of the complicated and dense band structure due to band folding, and the fact that the resulting flat bands are very close to neighboring bands[16,18] (**Fig. 4c**). For relatively large twist angles, experiments can also be challenging. One example of this is the topological corner states hosted by the SE even TBG[88], and such a corner state was numerically predicted to be located in an extremely narrow band gap in twisted photonic bilayer graphene, making its observation extremely difficult in experiments[18]. Additionally, in contrast to electronic systems, it is relatively easy to engineer flat bands or higher-order topological insulators using monolayer photonic and phononic crystals[89,90]. Therefore, it is crucial to elucidate the benefit of the twist-induced flat bands or topological corner modes in classical wave systems. While Moiré meta-crystals are tunable in nature due to their twist degree of freedom, and the results produced from classical wave systems can advance the research on TBG or twistronics in general by informing the discovery of new quantum materials, future work directions could leverage the engineering of flat bands and higher-order band topology for practical functionalities such as robust dynamic energy trapping via the twist, which could benefit the fields of nonlinear photonics and optomechanics.

The second direction entails a broader scope and it seeks to expand the field of artificial photonic and phononic meta-crystals by taking inspiration from TBG, in that the twist degree of freedom and interlayer coupling, or simply the Moiré pattern are harnessed to give rise to new design paradigms of classical wave devices. This line of research often leads to results that represent a significant departure from the TBG, in that these results find no counterparts in TBG. Earliest works that took a major step in this direction include the one that studied the twist between two layers of α-MoO$_3$[22] as a means to tailor the equal frequency contour, and the work that studied the localization and delocalization of light in photonic Moiré lattices[45]. Another example on this front is the bilayer valley Hall effect in twisted bilayer phononic graphene stemmed from the synergy between ultra-strong interlayer coupling and SE-even symmetry[87], which has not been observed in TBG owing to the intrinsically weak interlayer coupling. While significant progress has been made in the development of Moiré meta-crystals, there is still a need to explore their integration into functional devices that can leverage twist and interlayer coupling for precise wave control. One potential application involves the use of Moiré patterns in acousto-fluidics to create customizable fluid streaming patterns for the manipulation and trapping of microparticles. To achieve this goal, further research is required to investigate the physics and application of Moiré phononic meta-crystals in liquids. Currently, the majority of research in this field is focused on airborne sound and stress waves.

Going forward, there is a plethora of directions that can be explored to bring the field of Moiré photonic and phononic meta-crystals to the next phase. For example, non-Hermicity can be added into the equation to enrich the physics of Moiré photonic and phononic meta-crystals, where the interplay between loss and gain can be further complemented by twist and interlayer coupling. It is noted that, two recent papers have theoretically studied PT-symmetric AA- and AB-stack bilayer photonic graphene[91,92], and showed that PT symmetry induces band alteration in the vicinity of the Dirac point[92], and may lead to the existence of exceptional rings and exceptional concentric rings with particular topological features[91]. Another direction is to leverage the unique strength of PtCs and PnCs (or metasurfaces), where arbitrary 2D lattices other than the honeycomb lattice can be readily built, and their interaction with twist/interlayer coupling can be theoretically or even experimentally investigated. In this spirit, bilayer square lattice PtC and photonic Moiré pattern resulted from square lattices have been recently studied[45,69]. However, other lattices have been largely unexplored in classical waves such as Kagome lattice. Additionally, while in electronic materials, the nearest interlayer hopping is naturally the strongest, photonic and phononic meta-crystals can be a robust platform to engineer interlayer coupling, where next-nearest neighbor interlayer hopping can be made stronger than nearest interlayer hopping. A recent paper theoretically demonstrated that long-range hopping stronger than the nearest neighbor hopping can extend the topological order to a new topological class, giving rise to a greater number of topologically protected states in a 2D monolayer crystal[93]. We expect that new physics can be similarly uncovered in the bilayer system.

Finally, nonlinear optical responses of the TBG have been investigated to demonstrate higher-order harmonic responses that are absent in monolayer or conventional bilayer graphene, which spawned the field of optotwistronics[33,94–96]. Nonlinear optical waves have also been extensively studied in PtC, leading to applications towards reduced-size multifunctional control of light, photonic circuits for optical communication, and multi-photon absorption. In mechanical waves, nonlinear dynamical behaviors were also studied in PnCs to achieve subwavelength wave control[97], acoustic nonreciprocity[98,99], soft material lattices for nonlinear wave control[100], and architected lattices for solitons manipulation[100–102]. However, nonlinear dynamic responses of twisted bilayer photonic and phononic meta-crystals has yet to be explored. By incorporating the twist degree of freedom in conjunction with interlayer coupling, photonic and phononic meta-crystals can achieve a whole new level of capability with highly customizable nonlinear dynamic behavior, which has the potential to revolutionize photonics and acoustics, leading to significant technological breakthroughs.